\documentclass[a4paper,11pt]{article}
\pdfoutput=1 % if your are submitting a pdflatex (i.e. if you have
\usepackage{jcappub} % for details on the use of the package, please
\usepackage[T1]{fontenc} % if needed
\usepackage[english]{babel}
\usepackage{algorithm}
\usepackage{algorithmic}
\usepackage{xcolor}
\usepackage{mathdots}
\usepackage{subcaption}
\usepackage{layout}
\usepackage{printlen}
\usepackage{hyperref}
\usepackage{comment}
\usepackage[colorinlistoftodos]{todonotes}
\usepackage[normalem]{ulem}

\newcommand{\Planck}{\textit{Planck}\ }
\newcommand{\CosmoMC}{\texttt{CosmoMC}\ }
\newcommand{\LCDM}{$\Lambda$CDM\ }

\title{\boldmath Optimising Inflationary Features the Bayesian Way}

\author{Jan Hamann}
\author{and Julius Wons}
\affiliation{Sydney Consortium for Particle Physics and Cosmology, School of Physics, The University
of New South Wales, Sydney NSW 2052, Australia}

\emailAdd{jan.hamann@unsw.edu.au}
\emailAdd{j.wons@unsw.edu.au}

\abstract{Modern cosmological data demand modern data analysis techniques.
We introduce \emph{BayOp}, a new likelihood sampling and maximisation method which is based on the Bayesian Optimisation algorithm and learns a function instead of randomly sampling from it.  We apply BayOp to analyse \Planck data for traces of inflationary features models with global periodic modulations of the primordial power spectrum.  While we do not find any new evidence for features, we demonstrate that BayOp provides an extremely efficient way of sampling likelihoods over low-to-moderate-dimensional parameter spaces, even for very complex likelihood landscapes.}

\begin{document}
\begin{flushright}
\large \tt{CPPC-2021-10}
\end{flushright}

\maketitle
\flushbottom

\section{Introduction}
\label{sec:intro}
It has been over 40 years since the concept of cosmic inflation was introduced~\cite{Starobinsky:1979ty,Sato:1980yn,Guth:1980zm,Kazanas:1980tx}.  At first, the motivation was to explain fine-tuning and causality issues encountered in a radiation-dominated early Universe, but it was soon realised that inflation also provides a mechanism to generate primordial inhomogeneities and anisotropies that form the seeds of structure formation during a later era of the Universe~\cite{Mukhanov:1981xt}.
In the meantime, cosmological observations have all but confirmed the predictions of inflation, indicating that we live in a spatially flat Universe whose primordial fluctuations are to a very good approximation Gaussian, adiabatic and scalar and can be described by an almost scale-invariant power spectrum~\cite{Planck:2018vyg}.  This has become implicitly incorporated in the current minimal cosmological model (base $\Lambda$CDM) by taking the initial curvature perturbations to follow a power-law spectrum $\mathcal{P_R}(k) \propto A_\mathrm{s} k^{n_\mathrm{s}-1}$, which in itself is a fairly generic prediction of slow-roll inflation.  

On the quest to pin down the precise nature of the field(s) that drove inflation, this genericity presents a bit of a challenge.  While certain classes of inflationary models have been ruled out by current data~\cite{Vennin:2015eaa,Akrami:2018odb}, having effectively only two well-measured quantities ($A_\mathrm{s}$ and $n_\mathrm{s}$) plus a few parameters with weak constraints or only upper limits (e.g., non-Gaussianity $f_\mathrm{nl}$ parameters, the running of the spectral index or the tensor-to-scalar ratio) does not allow us to decide with any degree of certainty which particular model is realised in our Universe.  Now, if the true model was indeed of the generic canonical single-field slow-roll kind, there would be little else one could do beyond holding out for tighter constraints from future observations.  But that is of course not a given, and the possibility remains that whatever physics governed the earliest moments of our Universe did leave behind more unique traces for us to observe today.

One example of such unique traces is inflationary \emph{features}, i.e., strong deviations from a smooth power-law behaviour in the primordial power spectrum~\cite{Slosar:2019gvt}.  Even if \emph{a priori} theoretical prejudice would prefer the simplicity of plain slow-roll scenarios, features are not an absurd proposition and can be generated in a wide variety of well-motivated physical scenarios (see ref.~\cite{Chluba_2015} for a review).  Importantly, in case of a conclusive detection of features, one would not only be able to rule out the generic scenario, but since features contain a high degree of informativity, a determination of their shape would typically allow an identification of the underlying physical mechanism.  When it comes to searching for features, the presently most constraining observable are the anisotropies of the cosmic microwave background (CMB) which provide very good sensitivity over a wavenumber range spanning roughly three orders of magnitude.\footnote{In the future, the CMB will likely be superseded in this role by observations of tracers of the Universe's large scale structure, due to their ability to provide a three-dimensional map of the fluctuations, resulting in potentially better spectral resolution if the survey volume is sufficiently large. See, e.g., Ref.~\cite{Xu:2016kwz} for a forecast of the sensitivity of 21~cm-observations to inflationary features.}  Previous analyses of CMB data searching for inflationary features include, e.g. refs.~\cite{WMAP:2003syu,Martin:2003sg,Covi:2006ci,Meerburg:2011gd,Meerburg:2013dla,Easther:2013kla,Achucarro:2014msa,Hazra:2016fkm,Obied:2017tpd,Braglia:2021ckn}, but no statistically significant detection has been made to date.

While for many cosmological models, parameter estimation with CMB data can rely on a very well-developed analysis machinery~\cite{Lewis:2002ah,Brinckmann:2018cvx}, typically based on Monte Carlo sampling techniques, its application to features models is somewhat less straightforward.  This is for two reasons: a longer computation time due to increased requirements for numerical accuracy, and a complicated, often multimodal likelihood structure which renders conventional methods quite inefficient. 
In this work, we propose to address this problem by \emph{learning} the likelihood function, rather than generating samples more or less blindly through a random process.  Our approach is based on a method known as Bayesian Optimisation (BayOp)~\cite{Mockus1975,frazier2018tutorial} that takes into account the information gained by all previous samples to inform a ``smart'' sampling strategy.  We demonstrate this method by applying it to \Planck temperature and polarisation data~\cite{Planck:2018nkj}.

The structure of the paper is as follows: In Section~\ref{sec:BayOp}, we review the theoretical and statistical background of the BayOp algorithm. Then, in Section~\ref{sec:BayOpCosmo} we discuss the implementation of the algorithm into \texttt{CosmoMC} and the different features models we will analyse.  We present the results of our BayOp analysis of \Planck data and compare them to previous findings in Section~\ref{sec:results} and conclude in Section~\ref{sec:conclusions}. Further details on the code can be found in Appendix~\ref{app:A}.

\section{Bayesian Optimisation \label{sec:BayOp}}
The ultimate objective of BayOp is to extremise a real function $f(x)$, defined over a domain $\mathfrak{P} \subseteq \mathbb{R}^N$, which we will refer to as \textit{parameter space} in the following.  BayOp is very well suited for problems where
\begin{itemize}
    \item {the structure of $f$ is initially unknown (so-called black-box functions) and thus methods that exploit features such as linearity or convexity cannot be applied,}
    \item{$f$ has multiple local extrema, or}
    \item{$f$ is (literally or computationally) expensive to evaluate and efficiency is desired.}
\end{itemize}
In addition, BayOp can be implemented in such a way that it not only finds the function's extremum, but also returns an estimate of $f$ over the entirety of parameter space and at the same time quantifies the uncertainty of this estimate.

In this paper, we identify $f$ as the \Planck likelihood functions for various inflationary features models, and correspondingly $x$ is an element of the space of feature parameters.  The likelihoods in these models tend to have a highly nontrivial structure with many local maxima in some parameter directions. Furthermore, they also take considerably longer to compute than the likelihood for ``ordinary'' \LCDM fits, due to higher demands on the numerical precision in the calculation of CMB power spectra and the need to use the unbinned version of the \Planck likelihood.  In other words, all three points mentioned in the previous paragraph hold true for these scenarios, suggesting that BayOp will be a good choice for their analysis.

\begin{algorithm}
\caption{Bayesian Optimisation}
\label{algo:BayOp}
\begin{algorithmic}
\STATE generate initial set of function samples by evaluating the function at random points
\WHILE{acquisition function > threshold}
\STATE perform Gaussian Process Regression on samples to predict target function
\STATE evaluate acquisition function
\STATE compute function at the point that maximises the acquisition function 
\ENDWHILE
\end{algorithmic}
\end{algorithm}

The BayOp algorithm, as sketched in Algorithm~\ref{algo:BayOp}, is built around a loop of three basic steps.
Starting from a set $\mathbf{x}_D$ of samples in parameter space for which the corresponding function values $f(\mathbf{x}_D)$ are known, we make a prediction of $f$ conditional on the known samples over a grid $\mathbf{x}_P$ covering the entire parameter space, $f(\mathbf{x}_P | f(\mathbf{x}_D))$. The prediction is done by Gaussian Process Regression (GPR), which returns expectation values $\mu(\mathbf{x}_P)$ and associated variances $\sigma^2(\mathbf{x}_P)$ at all grid points. 

In the second step, we identify the grid point at which to perform the next computation of the function by evaluating a suitable acquisition function at all grid points and identify the grid point $x_P^\mathrm{max}$ which maximises the acquisition function. 
In the final step, we compute $f(x_P^\mathrm{max})$, add it to our set of samples and remove $x_P^\mathrm{max}$ from the grid.

With each iteration of this loop, we update our knowledge of $f$ and the quality of our prediction improves.  The algorithm continues until the maximum of the acquisition function drops below a pre-determined threshold value (which can be related to an accuracy goal for the global maximum of $f$). Let us now delve a bit deeper into the details of the algorithm's steps.

\subsection{Gaussian Process Regression}
In the first step of the algorithm, the goal is to predict the behaviour of the function $f$ without actually evaluating it, but using instead the information contained in a set of known function values -- the {\it data} -- making this essentially an interpolation/extrapolation or {\it regression} problem.  Rather than assuming some fixed functional form such as linear functions or splines for the regression, the basis of GPR is a Gaussian process, in which all function values are interpreted as correlated Gaussian random variables~\cite{williams2006gaussian,Aghamousa:2017uqe}, with the correlation between function values at points $x_i$, $x_j$ encoded in the kernel function $k(x_i,x_j)$:
\begin{equation}
    {\rm cov}\big(f(x_i),f(x_j)\big) \equiv k(x_i,x_j).
\end{equation}
The result of the regression thus becomes a distribution over random realisations of the Gaussian process conditional on the data.  By virtue of the assumed Gaussianity, this distribution is fully characterised by its mean and covariance.

The kernel function plays a crucial role in determining the properties of the Gaussian process, and appropriate choices of $k(x_i,x_j)$ can capture certain features of the regression (e.g., periodicity or linearity), see ref.~\cite{duvenaud2014automatic} for an in-depth discussion.   
In the absence of any prior knowledge about $f$ and motivated by a desire for flexibility, we will in this work employ a Gaussian kernel,
\begin{equation}\label{eq:kernel}
    k(x_i,x_j) = \sigma_f^2 \, \exp\left(- \frac{(x_i-x_j)^2}{2\,l^2}\right)\,,
\end{equation}
which yields infinitely differentiable Gaussian processes with local variations at the scale of the correlation length $l$ and a typical amplitude of order the prior width $\sigma^2_f$.  The parameters of the kernel are known as hyperparameters.  At first glance, it might seem that having to input their values by hand would introduce a great degree of arbitrariness and model-dependence if we are dealing with an unknown function $f$, which is something this approach was meant to avoid.  However, as we shall see in Section~\ref{sec:hypers} below, this issue can be avoided since it is actually possible to let the data select the most suitable values of the hyperparameters. 

With the kernel function defined above and two sets of elements of parameter space, \mbox{$\mathbf{x} = \{ x_1, ..., x_n \}$} and $\mathbf{x}' = \{ x'_1, ..., x'_m \}$,
we define the kernel matrix $K$
\begin{equation}
%    K(\Vec{x},\Vec{x}\,')\equiv
    K(\mathbf{x},\mathbf{x}')\equiv
    \begin{pmatrix}
    k(x_1,x'_1) & \dots & k(x_1, x'_m)\\
\vdots & \ddots & \vdots\\
k(x_n,x'_1) &\dots & k(x_n,x'_m)
    \end{pmatrix}\,,
\end{equation}
which has dimension $n\times m$.

For a set of input data points $\mathbf{x}_D$ and a set of prediction points $\mathbf{x}_P$, the Gaussian process follows a joint distribution function given by
\begin{equation}
    \begin{pmatrix}
    f(\mathbf{x}_D) \\
    f(\mathbf{x}_P)
    \end{pmatrix}
    =\mathcal{N}\left(\mu_0,    \begin{pmatrix}
    K_{DD} & K_{DP}\\
    K_{DP}^T & K_{PP}
    \end{pmatrix}\right)\,,
\end{equation}
where $\mathcal{N}(\mu_0,\Sigma_0)$ is the multivariate normal distribution with mean $\mu_0$ and covariance $\Sigma_0,$\footnote{
Generally, the mean of the Gaussian process could be an arbitrary function of $x$.  In practice, it is often taken to be a constant, e.g., the mean of the data $f(\mathbf{x_D})$.  In the following we will assume that $\mu_0 = 0$.  This can be done without loss of generality, since one can simply transform the function instead: a GPR for $f(x)$ with $\mu_0 = \mu_0(x)$ is equivalent to a GPR for $\tilde{f}(x) \equiv f(x) - \mu_0(x)$ with $\tilde{\mu}_0 = 0$.}
and the covariance is composed of the kernel matrices
\begin{equation}\label{eq:Kernels}
    K_{DD}\equiv K(\mathbf{x}_D,\mathbf{x}_D)\,, \qquad K_{DP}\equiv K(\mathbf{x}_D,\mathbf{x}_P)\,, \qquad K_{PP}\equiv K(\mathbf{x}_P,\mathbf{x}_P)\,.
\end{equation}
The conditional distribution $f(\mathbf{x}_P | f(\mathbf{x}_D))$ is a multivariate Gaussian whose expectation values and covariance are given in terms of the data and the kernel matrices by~\cite{williams2006gaussian}:
\begin{align}\label{eq:mu}
    \mu(\mathbf{x}_P) &= K_{DD} K_{DP}^{-1}f(\mathbf{x}_D),\\
    \mathrm{cov}(\mathbf{x}_{P},\mathbf{x}_P) &= K_{PP} - K_{DP}^TK_{DD}^{-1}K_{DP}\,.
\end{align}    
The expectation values $\mu(\mathbf{x}_P)$ are the result of our prediction and we can quantify the individual uncertainty of the regression for each element of $\mathbf{x}_P$ using the variance
\begin{equation}    
    \sigma^2(\mathbf{x}_P) = {\rm diag(cov(\mathbf{x}_P,\mathbf{x}_P))}= \sigma_f^2- \mathrm{diag}\left(K_P^TK_D^{-1}K_P\right)\,. \label{eq:variance}
\end{equation}

In the limit of large distances between a prediction point and the nearest data point \mbox{$|x_P - x_D| \gg l$}, the prediction $f(x_P)$ becomes uncorrelated with the data (i.e., the data do not provide any information), its uncertainty $\sigma$ is the prior width of the Gaussian process $\sigma_f$ and $\mu \rightarrow \mu_0$.
Conversely, the closer $x_P$ is to the nearest data point $x_D$, the more information we can gain about $f(x_P)$ and the smaller the uncertainty, with $\sigma = 0$ in the limit of coincidence $x_P = x_D$.  We illustrate this behaviour in the middle panel of figure~\ref{fig:correlation}.

\subsection{Determining the hyperparameters of the kernel \label{sec:hypers}}
%Our choice of t
The kernel in eq.~\ref{eq:kernel} includes two hyperparameters: the correlation length $l$ and the 
prior width $\sigma^2_f$.  The ability of the Gaussian process to emulate the underlying function 
%(i.e., the quality of the regression) 
depends on these parameters.

\begin{figure}[t]
    \centering
    \includegraphics[width=0.88\textwidth]{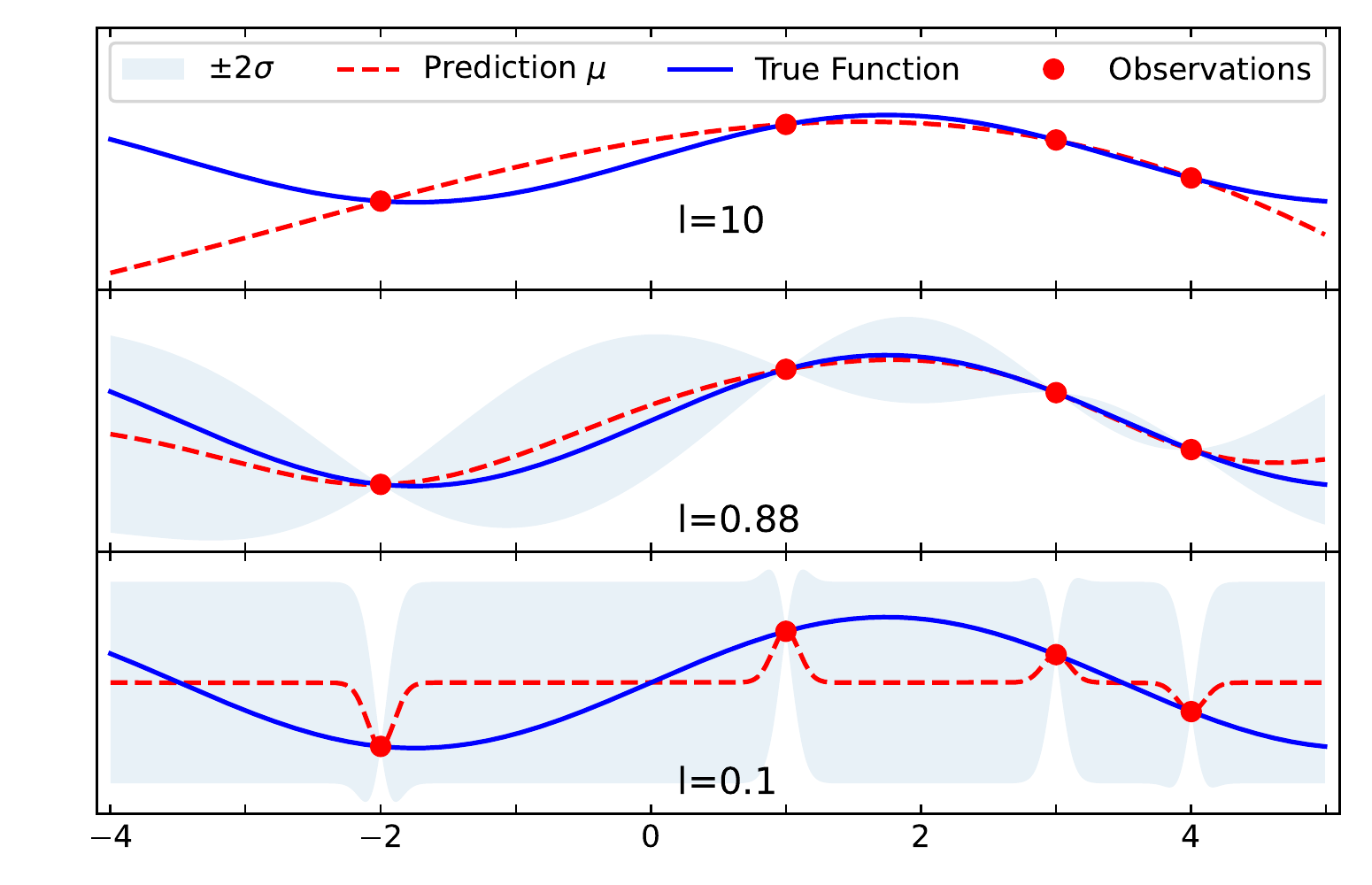}
    \caption{Gaussian Process Regression on four data points taken from a sine function for different correlation lengths. The dashed red line is the expectation value of the Gaussian process $\mu(x)$, the shaded band denotes the $\pm 2 \sigma$ uncertainty, and the solid blue line is the true function. In the middle, the correlation length is determined by maximising the marginal likelihood, resulting in $l=0.88$. At the top, we chose an overly large correlation length of $l=10$. At the bottom, the correlation length is too small with $l=0.1$. \label{fig:correlation}}
\end{figure}

As an example, we demonstrate in figure~\ref{fig:correlation} how the prediction depends on the correlation length $l$. Note that the prior width $\sigma^2_f$, quantifying the overall amplitude of the uncertainty, is chosen to be the same in all examples.  In the top panel, a too-large correlation length is chosen. The predicted function only varies slowly and is too ``stiff'' to match the underlying oscillation pattern correctly. Additionally, due to the strong correlation, the uncertainty of the prediction remains very small, even far away from data points. In the bottom panel, a too-short correlation length is chosen. The prediction returns to the mean of zero a few correlation lengths away from the data. Most of the parameter space is uncorrelated, which means it has zero mean and maximal uncertainty. The true function is contained well inside the uncertainty band, but the predictive power is very low.  Clearly, for our example, a more appropriate choice of $l$ must lie somewhere between these two cases.

In general, the quality of the regression can be quantified in terms of the logarithm of the marginal likelihood of the data~\cite{williams2006gaussian},
\begin{equation}\label{eq:MLE}
    \log p(f(\mathbf{x}_D)|\mathbf{x}_D) = -f(\mathbf{x}_D)^T K_{DD}^{-1}f(\mathbf{x}_D) - \frac{1}{2}\ln|K_{DD}|-\frac{n}{2}\ln2\pi \,.
\end{equation}
The right hand side depends on the kernel function, and thus the hyperparameters.  Hence, we can find the optimal choice of hyperparameters for a given set of data by finding the combination which maximises $\log p(f(\mathbf{x}_D)|\mathbf{x}_D)$, noting that there is no dependence on the prediction points $\mathbf{x}_P$.
 We show the result of this optimal choice for $l$ in our example in the middle panel of figure~\ref{fig:correlation}.  The prediction is very close to the underlying function, and while the uncertainty grows with distance to the data points, the true function is always well inside the 2$\sigma$ uncertainty bands.

While in this section we have so far implicitly assumed a one-dimensional parameter space with $x \in \mathbb{R}$, GPR can straightforwardly be generalised to the $N$-dimensional case \mbox{$x = (x^{(1)},\dots,x^{(N)}) \in \mathbb{R}^N$}: one merely needs to define a new kernel function.  In the rest of this work, we will use a factorised Gaussian kernel with a separate correlation length for each parameter direction:
\begin{equation}\label{eq:kernel2D}
        k(x_i,x_j) = \sigma_f^2 \,\prod_{n=1}^N\exp\left(- \frac{\left(x_{i}^{(n)}-x_{j}^{(n)}\right)^2}{2 \, l_n^2}\right)\,.
\end{equation}
This leaves us with $N+1$ hyperparameters altogether, whose optimal values will be determined by finding global maximum of $\log p(f(\mathbf{x}_D)|\mathbf{x}_D)$.

\subsection{Expected Improvement}
Having obtained a prediction of the target function via Gaussian Process Regression, the next step is to identify the most promising point to sample next.  To this effect, we can define an acquisition function that guides our search for the global maximum and is computationally cheap to optimise. 

Ideally, the acquisition function will strike a good balance between recommending new samples to be taken from regions of parameter space close to known large function values with corresponding small uncertainties ({\it exploitation}) and regions far away from already collected samples, which are poorly known and have large uncertainties associated with them ({\it exploration}).
Various choices of acquisition functions have been proposed in the literature~\cite{frazier2018tutorial,brochu2010tutorial}.  In this paper, we implement Expected Improvement (EI) as our acquisition function, as it is widely used and has been shown to perform well in most settings \cite{lizotte2008practical}.

If we denote the current largest sampled function value as $f_+$ and the expectation value of the prediction at a given point $x$ as $\mu(x)$, with corresponding uncertainty $\sigma(x)$, the Expected Improvement is defined as
\begin{equation}\label{eq:EI}
    \textrm{EI}(x)=\
    \left( \mu(x) - f_+ -\xi\right)\Phi(Z)+\sigma(x)\phi(Z)
\end{equation}
with
\begin{equation}
    Z =
    \begin{cases}
    \frac{\mu(x)-f_+-\xi}{\sigma(x)} &\textrm{if} \qquad \sigma(x)> 0
    \,,\hspace{5mm}\\
    \hspace{7mm} 0 &\textrm{if} \qquad \sigma(x) = 0
    \,,
    \end{cases}
\end{equation}
where $\phi(x)$ is the probability density function of the standard normal distribution, and $\Phi(x)$ is the cumulative distribution function of the standard normal distribution.

As the name suggests, the EI can be derived by taking the expectation value of the improvement 
\begin{equation}
I(x)={\rm max}(0,\,\mu(x)-f_+)\,.
\end{equation}
A non-zero expectation value can be caused by two effects. First, if the prediction $\mu(x)$ is larger than the current best point $f_+$, i.e., the improvement itself $I(x)$ is positive. This case is mainly driven by the first term in eq.~\eqref{eq:EI} and exploits our knowledge of the function. 
Even if $\mu(x)<f_+$ when the improvement $I(x)$ is vanishing, the EI can be non-zero. In this instance, the first term in eq.~\eqref{eq:EI} is negative, but as long as the second term dominates over the first term for large uncertainties $\sigma(x)$, the EI will be positive. This second case leads to the exploration of the parameter space.

The parameter $\xi$ can be tuned to shift the balance between exploration and exploitation. For $\xi>0$, the first term in eq.~(\ref{eq:EI}) is suppressed and more weight is given to exploration; conversely, $\xi<0$ enhances the first term and encourages exploitation.  Ref.~\cite{lizotte2008practical} notes that a setting of $\xi=0.01-0.1$ does not perform significantly worse than other choices in many settings. However, we found that for our particular problem, $\xi=0$ is a very good choice.  The likelihood function's complex structure results in a natural tendency for the algorithm to explore a lot, and no further boost to exploration is necessary.

Note that if it were our goal to learn the function in the most efficient way rather than finding the global maximum, one would always sample at the point of greatest uncertainty (corresponding to the limit $\xi \gg \max (|\mu(x) - f_+|)$ in eq.~\eqref{eq:EI}).

\begin{figure}[t]
    \centering
    \includegraphics[width=0.98\textwidth]{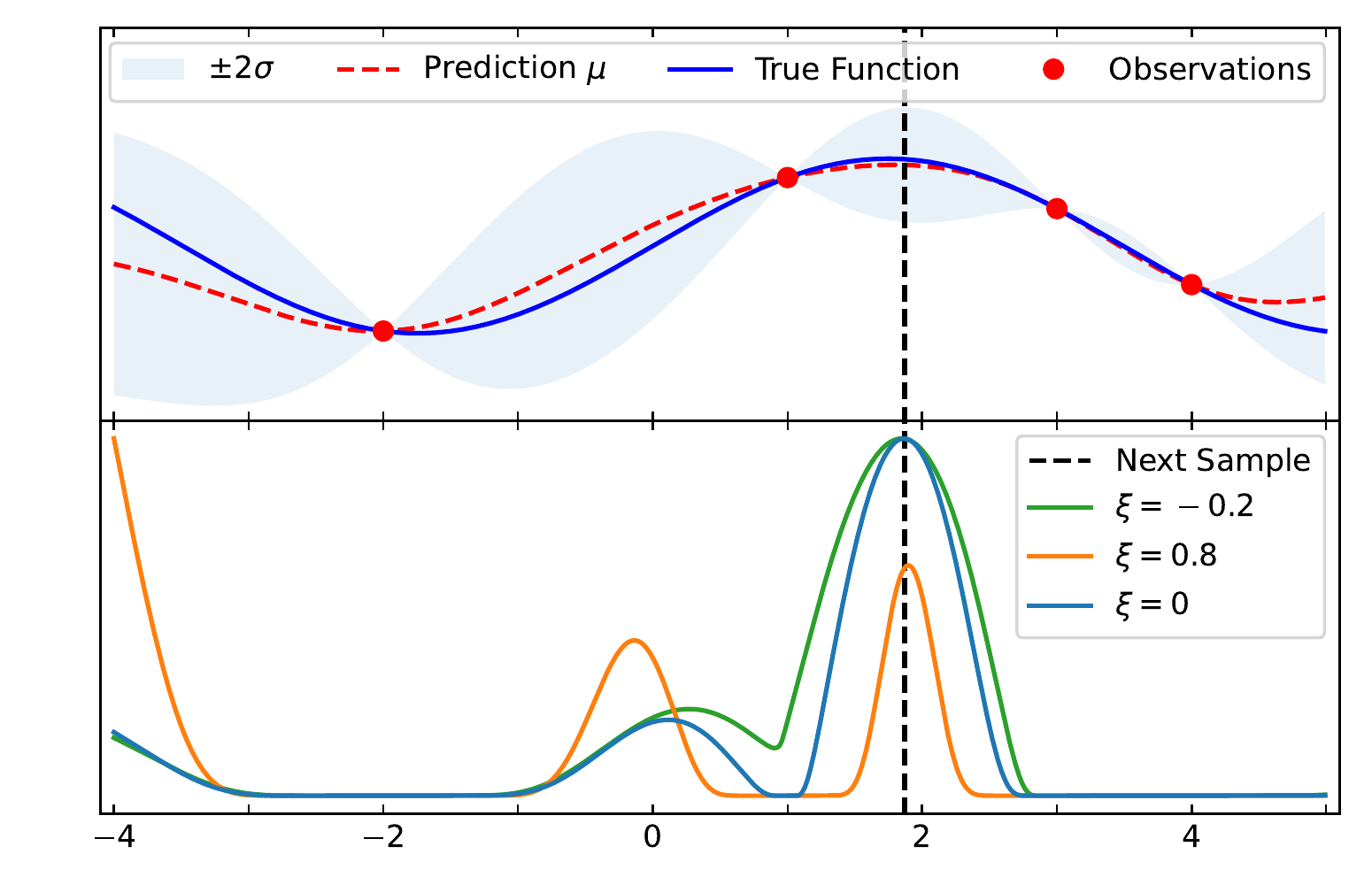}
    \caption{\label{fig:BayOp} The Gaussian Process is the same as in the middle panel of Fig~\ref{fig:correlation}. The corresponding expected improvement is plotted at the bottom for different values of $\xi$. The functions have been rescaled to yield the same value at the maximum. The dashed line marks the maximum of the expected improvement, i.e., the point at which the next sample will be taken.}
\end{figure}

In figure~\ref{fig:BayOp}, we plot the GPR for our previous sine function example along with the corresponding EI.  Regions where the upper edge of the uncertainty band exceeds the current maximum sample ($\mu+2\sigma>f_+$) make good candidates for drawing the next sample and show up as peaks in the EI, with the most promising point given by the global maximum of the EI.

We also show the EI for different values of $\xi$ to illustrate the effect of this parameter. The orange line with $\xi=0.8$ has an increased focus on exploration, and the next sample point would be at $x = -4$ where a large uncertainty is combined with a modestly high mean. For negative $\xi$, the regions with a promising expectation value are given more weight instead.

After calculating the EI for all elements of $\mathbf{x}_P$, we identify the point $x_P^\mathrm{max}$ which maximises the EI, evaluate $f(x_P^\mathrm{max})$, remove $x_P^\mathrm{max}$ from $\mathbf{x}_P$ (in order to avoid it being selected more than once), and add it to the set of data points $\mathrm{x}_D$, whereupon the whole process is repeated with the updated $\mathrm{x}_D$ and $\mathrm{x}_P$.

\subsection{Stopping the algorithm}
As we cycle through the algorithm and keep generating new data points,  there is a general tendency for the EI to decrease as more and more of the parameter space gets explored and local maxima exploited.  While it may occasionally increase if an exploration step happens to find an unexpectedly high function value, possibly associated with a previously unknown (local) maximum, this becomes increasingly unlikely the longer the algorithm runs.  It thus appears reasonable to exploit this behaviour of the EI maximum to define a stopping criterion for the sampling loop, namely when the highest EI drops below a threshold value $\epsilon_\mathrm{EI}$ (see also, e.g., ref.~\cite{nguyen2017regret}).

Since the EI roughly indicates the expectation by how much the next sample could exceed the current maximal value of $f$, the threshold
$\epsilon_\mathrm{EI}$ also represents an approximate measure of the accuracy of our final estimate of the function's maximum.

We conservatively take $\epsilon_\mathrm{EI}$ to be a few orders of magnitude below the desired accuracy of $\sim 0.1$ and choose $\epsilon_\mathrm{EI} = 10^{-5}$.  In any case, in the problems considered here, the EI usually drops very quickly away from the peaks and we do not find a strong dependence of the algorithm's runtime for values of $\epsilon_\mathrm{EI}$ in the range $10^{-5}$--$10^{-15}$. We describe the implementation of the threshold and its impact on the overall runtime in more detail in appendix~\ref{app:remove}.

\section{Bayesian Optimisation on cosmological data \label{sec:BayOpCosmo}}
Having introduced the BayOp algorithm, let us now turn to its application to a concrete physical scenario: the search for primordial power spectrum features in CMB data.

\subsection{Features in the primordial power spectrum \label{sec:features}}
In the well-established base \LCDM model of cosmology, the initial inhomogeneities are characterised by a primordial power spectrum of scalar perturbations that follows a simple power law:
\begin{equation}
    \mathcal{P}_\mathcal{R}^0(k)=\mathcal{A}_s \left(\frac{k}{k_*}\right)^{n_s-1}\,,
\end{equation}
where $\mathcal{A}_\mathrm{s}$ is the amplitude of the power spectrum at the pivot scale $k_*=0.05\,{\rm Mpc}^{-1}$ and $n_\mathrm{s}$ is the spectral index. Such a primordial power spectrum generically arises in standard single field slow-roll models of inflation with canonical initial conditions. 

In case any of these conditions are violated, $\mathcal{P}_\mathcal{R}(k)$ may significantly deviate from a power-law behaviour and exhibit various kinds of \emph{features}.  In particular, various types of oscillatory modulations of the power spectrum are a very typical signature~\cite{Chluba_2015}.

In our attempt to determine whether or not \Planck CMB data show evidence for such oscillatory features, we will follow a ``top-down'' approach (e.g.,~\cite{WMAP:2003syu,Martin:2003sg}), 
by comparing the fit of a power-law spectrum to the fits of modulated power spectrum templates of form
\begin{equation}\label{eq:parametrisation}
    \mathcal{P}^X_\mathcal{R}(k)=\mathcal{P}^0_\mathcal{R}(k)[1+\mathcal{A}_X \cos(\omega_X \Xi_X(k)+\varphi_X)]\,,
\end{equation}
where $\mathcal{P}^0_\mathcal{R}$ is the standard \LCDM power-law power spectrum and $X\in \{\textrm{log, lin, rf, psc}\}$. The primordial power spectra of the models we consider are given by
\begin{align}\label{eq:models1}
    \Xi_{\rm log}&\equiv\ln \frac{k}{k_*}\,, \qquad
    \Xi_{\rm rf}\equiv\ln  \frac{k}{k_*} \left(1+\alpha_{\rm rf} \ln  \frac{k}{k_*}\right)\,, \qquad \\ \label{eq:models2}
    \Xi_{\rm lin}&\equiv \frac{k}{k_*}\,, \qquad 
    \Xi_{\rm psc}\equiv p\left(\frac{k}{k_r}\right)^{\frac{1}{p}}\,,
\end{align}
with a pivot scale $k_* = 0.05~\mathrm{Mpc}^{-1}$.
Some of these, namely the linear (lin), logarithmic (log) and logarithmic oscillation with running frequency (rf) models have previously been studied on the same data with different methods in ref.~\cite{Akrami:2018odb} (PI2018). Therefore, we can use the PI2018 results as direct comparison to validate our approach.

In addition, we analyse primordial standard clock models (PSC) \cite{Chen:2014cwa,Chen:2015lza}. Modulations of this type arise if additional heavy fields are present and dynamically relevant during the generation of the primordial fluctuations. The parameter $p$ is related to the behaviour of the scale factor when the corresponding scales exit the horizon via $a(t)\sim t^p$ and can thus reveal extra information regarding the dynamics of the earliest moments of the Universe. 

Also, if $p$ is determined with sufficient accuracy, the degeneracy between a power spectrum with PSC in an inflationary fast-expanding universe and a contracting universe~\cite{Wands_1999} can be lifted, provided that non-trivial interactions between inflaton and the additional field are negligible~\cite{Domenech:2018bnf}. PSC models have been suggested as an explanation of the lensing anomaly in the CMB \cite{Aghanim:2018eyx,Domenech:2019cyh,Domenech:2020qay}. This approach will be tested by future E-polarisation data.

\subsection{Implementation into CosmoMC}
The analysis software used in this work is based on the June 2021 version of \texttt{CosmoMC}~\cite{Lewis:2002ah} and our own \texttt{FORTRAN} BayOp module, which we implement into \texttt{CosmoMC} as a new sampling option besides the regular choices (Metropolis-Hastings, etc.).
We add the features models of Section~\ref{sec:features} and their corresponding parameters to \texttt{CAMB}.  
Note that in its default settings, \texttt{CAMB} is optimised for standard cosmologies, however, if we demand sufficient numerical accuracy for primordial power spectra with high-frequency oscillations, we need to (i) calculate the angular power spectra $\mathcal{C}_\ell$ for every single multipole $\ell$ rather than interpolating, and, (ii) increase the wavenumber sampling in the numerical integration over the source function.
These changes considerably increase the runtime of \texttt{CAMB} such that the calculation of the $\mathcal{C}_\ell$ and evaluation of the likelihood for a single cosmology takes approximately 35s on a single core compared to 5s runtime with the default settings.
Likewise, when it comes to the \textit{Planck} high-$\ell$ likelihood, the default version is binned in $\ell$ and therefore not suited to detecting high-frequency features.  Instead, we use the the final release of the unbinned \Planck TTTEEE likelihood 
%\texttt{Plik bin1}
at $\ell \geq 30$, combined with the latest version of the  \texttt{commander\_dx12\_v3\_2\_29} and \texttt{simall\_100x143\_offlike5\_EE\_Aplanck\_B} likelihoods for low $\ell$ temperature and E-polarisation data, respectively.

Before starting the BayOp analysis, we determine the best-fit parameter values and the best-fit likelihood $\mathcal{L}^0_\mathrm{max}$ of the base \LCDM Model with a power-law primordial spectrum using \CosmoMC's minimisation algorithm.  
This serves two purposes.  Firstly, it establishes a baseline likelihood value to compare the features models against and allows us to define an effective 
\begin{equation}
    \Delta \chi^2 \equiv  2 (\ln \mathcal{L} - \ln \mathcal{L}^0_\mathrm{max})
\end{equation}
to quantify how much a feature improves ($\Delta \chi^2 < 0$) or deteriorates ($\Delta \chi^2 > 0$) the fit to the data.
Secondly, in order to keep the dimensionality of our problem manageable, we restrict ourselves to an analysis of the space of features parameters only and keep both the base \LCDM and the \Planck nuisance parameters fixed to their power-law spectrum best-fit values.
Let us point out that this restriction strictly underestimates the improvement in $\Delta \chi^2$ that features models can achieve.  However, with the possible exception of very low frequency modulations of the primordial spectrum and a narrow frequency range in the linear oscillation model which could mimic the acoustic oscillations of the CMB, the effects on the CMB spectra of the global oscillation models under consideration here are generally quite distinct from the effects of varying other parameters.  This means that degeneracies between the features parameters and the \LCDM/nuisance parameters will be largely negligible and therefore not strongly impact our estimates of $\Delta \chi^2$.

We cover the $N$-dimensional space of features parameters with a grid of $n_\mathrm{grid}^{0}$ points initially.  As new data samples are generated and subsequently removed from the initial grid, the sum of remaining grid points and data points in principle satisfies $n_\mathrm{grid}+n_\mathrm{data} = n_\mathrm{grid}^{{0}}$.  This number represents the main limitation of the BayOp approach, since each iteration of the algorithm requires us to solve an $n_\mathrm{grid} \times n_\mathrm{data} =
n_\mathrm{grid}^{0} (1-n_\mathrm{data}/n_\mathrm{grid}^{0})
\times n_\mathrm{grid}^{0} (1-n_\mathrm{grid}/n_\mathrm{grid}^{0})$ 
linear system.
Due to the exponential dependence of $n_\mathrm{grid}^{{0}}$ on $N$, and given a typical computer's memory, it is thus realistically only applicable to problems with a modest number of dimensions $N \lesssim 5\hbox{-}6$. 
Though once the shape of the function is reasonably well known, the performance of the algorithm can be improved by discarding grid points with very low expected improvement (refer to Appendix~\ref{app:remove} for a discussion of our implementation).

We find that before commencing with BayOp sampling, it is helpful to sample the likelihood function on a small number of randomly chosen grid points.  
This avoids issues with the algorithm initially oversampling the boundaries of parameter space, which can occur if we have essentially no information about the function in the interior of parameter space (see appendix~\ref{app:random} for further details).

With this initialisation done, we start the BayOp algorithm as described in section~\ref{sec:BayOp} and continue collecting data points until either the maximum EI falls below the threshold or there are no further undiscarded and unsampled grid points remaining. 

The computational overhead associated with each BayOp step (consisting of the optimisation of hyperparameters, GPR on the grid of prediction points, evaluation of EI, identifying the EI maximum) obviously varies throughout the analysis, depending for instance on the dimensionality of parameter space, the number of data points already sampled and the number of prediction points left.  For all examples considered in this work, the average time this takes is of $\mathcal{O}(1\,\mathrm{s})$, i.e., only a relatively small fraction of the time it takes to perform one likelihood calculation.  In other words, smart sampling can be had for cheap here!

\subsection{Differences between the PI2018 analysis and this work \label{sec:difference_PI2018}}
Before we present our results, let us discuss the differences between our analysis and the one in PI2018.

\begin{figure}[t]
    \centering
    \includegraphics[width=0.98\textwidth]{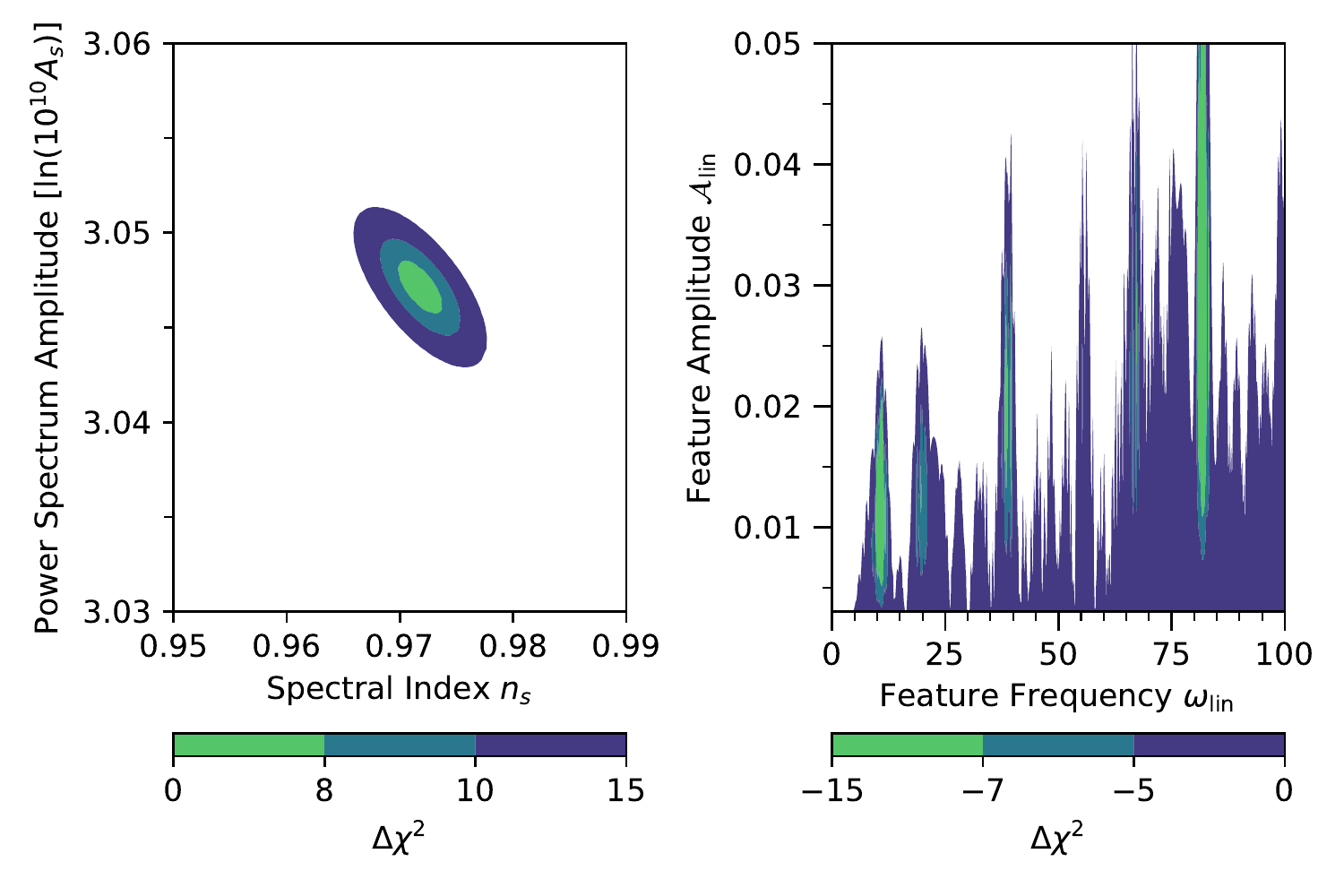}
    \caption{Two examples illustrating the structure of the posterior, profiled on different 2-dimensional slices of parameter space.  \textit{Left panel:} a smooth, convex and unimodal posterior; the color scale indicates the deterioration compared to the best fit point, white regions have $\Delta \chi^2 > 15$. \textit{Right panel:} a complicated, non-convex multimodal posterior;  the color scale indicates the improvement over the \LCDM best fit, with no improvement in white regions.
    \label{fig:convex}}
\end{figure}

First and foremost among these is of course the sampling method used.  The commonly employed Metropolis-Hastings algorithm~\cite{metropolishastings} works quite well on the simple unimodal posteriors often encountered in cosmological problems (see for instance the left panel of figure~\ref{fig:convex}), but tends to perform rather poorly in problems with multimodal posteriors such as the one we are considering here.  In features models, the posterior in the modulation frequency direction tends to have many local maxima (illustrated in the right panel of figure~\ref{fig:convex}).
For this reason, PI2018 used the nested sampling algorithm~\cite{Skilling:2006gxv} as implemented in \texttt{MultiNest}~\cite{Feroz:2008xx,Feroz:2013hea} -- which is able to identify local maxima via clustering analysis of samples and exploit them subsequently.  

Since new samples are drawn randomly from (a subset of) parameter space though, finding a local maximum in the first place to a certain extent relies on the luck of the draw, and there is always a chance that some maxima are missed, particularly for narrow peaks that contain only a small fraction of the posterior's total volume.  \texttt{MultiNest} nonetheless allows the reliable computation of marginalised posteriors and the Bayesian evidence as well as a reasonable estimate of profile posteriors and the global maximum.  It also does not scale too unfavourably with the dimensionality of the problem, and when combined with slice-based sampling (as in \texttt{PolyChord}~\cite{Handley:2015fda}), performs well for the $\mathcal{O}(30)$-dimensional parameter space involved in a typical full analysis of \Planck data.  All this comes, however, at the cost of requiring a large number of function evaluations.

As we shall see below, in comparison BayOp is more economical due to not having to rely on randomness (besides the initial sampling phase) and making use of the full information contained in the sampled points.  It results in a more precise determination of global maxima and a more reliable estimate of profile posteriors, including uncertainties. It also has the advantage that its performance does not require fine-tuning of parameter settings. It is not, however, applicable to problems with high dimensionality and does not yield marginalised posteriors, nor does it compute the Bayesian evidence.\footnote{At least not in our current implementation.  These quantities could in principle also be obtained via BayOp, but would require a different choice of acquisition function that prioritises learning the global shape of the function rather than maximisation.}  BayOp should thus be regarded as a complementary tool to Markov chain based samplers.

On top of this, there are differences in three additional aspects of the analysis:
firstly, PI2018 did include the non-nuisance LCDM parameters whereas we do not.  Our results thus strictly underestimate the true $\Delta \chi^2$, but as it turns out, this does not have actually a noticeable effect.  Secondly, rather than running BayOp on the entire parameter space at once, we take advantage of the easy parallelisability of BayOp and split parameter space in the frequency parameter direction into several non-overlapping sections.  This ensures both a better local fit of hyperparameters of the Gaussian process, as well as decreasing the chance of missing lesser peaks. We initially considered a range of choices for the number (and size) of bins and compared the code's performance for the different settings. Efficient choices of bin width can be related to the correlation length in the corresponding parameter direction (see~\ref{app:split} for further details). And finally, instead of a logarithmic scale, we use a linear scale in the feature frequency.  Since the peaks of the posterior in $\omega$ have approximately frequency-independent widths $\Delta \omega \sim 2$, a log scale in $\omega$ makes peaks at higher frequencies narrower and therefore easier to miss.

\section{Results \label{sec:results}}

In this section, we present the results of the BayOp analysis of the models introduced in section~\ref{sec:features}.  A summary of the best-fit parameters and $\Delta \chi^2$ for each model is given in~table~\ref{tab:chisq}.

\begin{table}[t]
    \centering
    \begin{tabular}{c|c|c|c||c|c|c}
        \hline \hline   &  Logarithm & Running Log  & Linear
                        &  Inflation       & Bounce       & Ekpyrotic\\ \hline
        $\Delta\chi^2$  & -10.6      & -12.2       & -12.5
                        & -13.2      & -11.8        & -13.8      \\ \hline
        $\mathcal{A}_X$ & 0.015      & 0.033        & 0.039     
                        & 0.07        & 0.013        & 0.305      \\
        $\omega_X$      & 18.2       & 59.8         & 81.7     
                        & 99.9       & 55.0         & 304.6      \\
        $\varphi_X/2\pi$& 0.08       & 0.86         & 0.64       
                        & 0.26       & 0.40         & 0.67      \\
        $\alpha_\mathrm{rf}$   & -          & 0.086        & -       
                        & -          & -            & -      \\
        $p$             & -          & -            & -       
                        & -          & 0.668        & 0.244      \\
        $k_r/{{\rm Mpc}^{-1}}$           & -          & -            & -       
                        & 0.069         & 0.168         & 0.0138
        
    \end{tabular}
    \caption{Best fit parameter values and $\Delta\chi^2$ relative to \LCDM for the different models of section~\ref{sec:features}.}
    \label{tab:chisq}
\end{table}

\subsection{Comparison between BayOp and PI2018}
%In the PI2018 analysis \cite{Akrami:2018odb}, the linear oscillations and the logarithmic oscillations with and without running frequency (Eq.~\eqref{eq:parametrisation}) were analysed using the MCMC algorithm in \CosmoMC as described in Sec. \ref{sec:difference_PI2018}. We compare their findings to the results derived using the BayOp sampler.
Since the PI2018 analysis was performed on a subset of the models considered here and involved the same likelihoods, their results can be used to validate our approach.

\begin{figure}[t]
    \centering
    \includegraphics[width=0.98\textwidth]{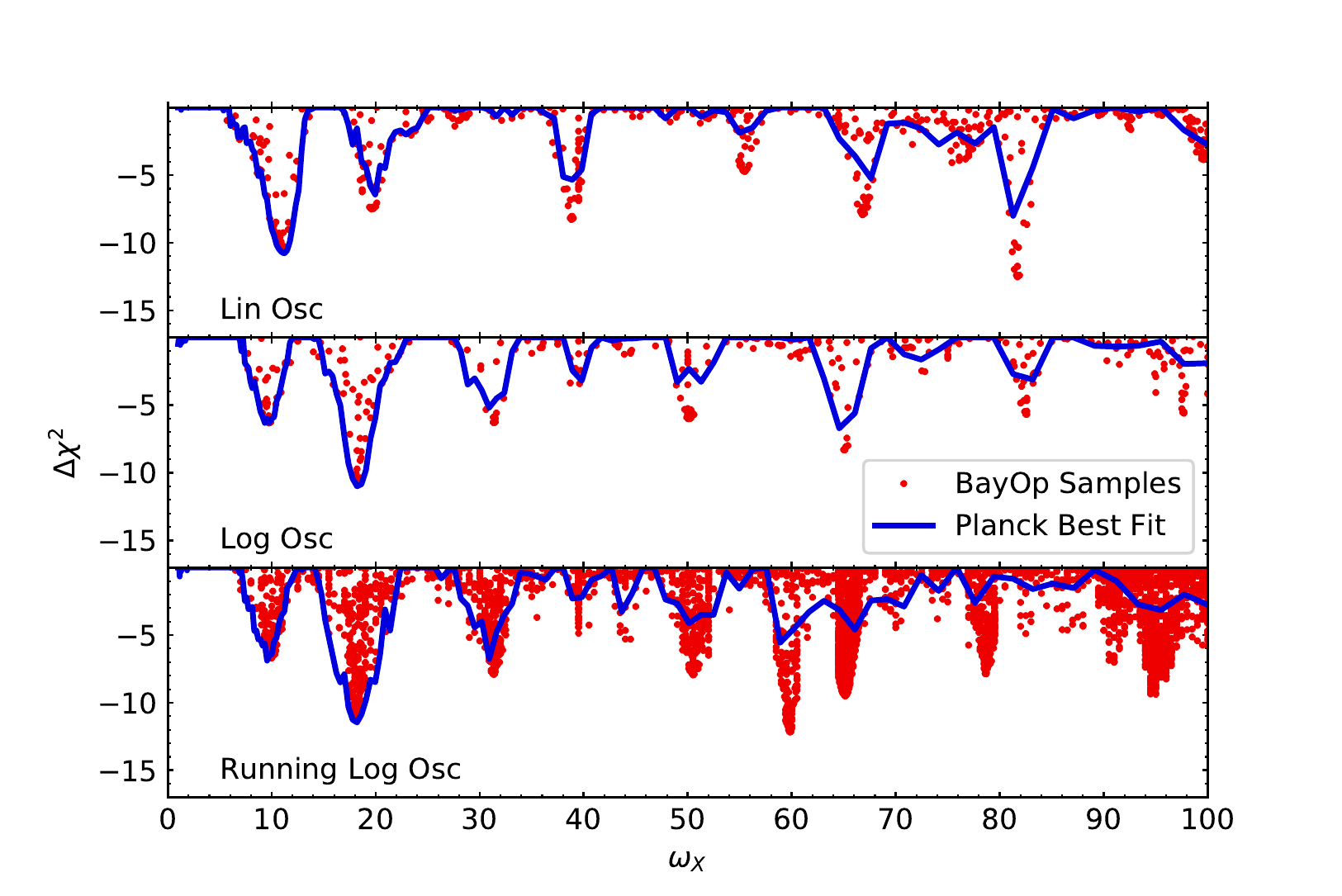}
    \caption{\label{fig:Planck_xisq}
    Profile likelihoods of the \Planck temperature+polarisation data in the frequency parameter direction for the linear oscillation (\textit{top}), logarithmic oscillation (\textit{centre}) and running logarithmic oscillation models (\textit{bottom}), using the parametrisation defined in eq.~\ref{eq:parametrisation}.  The blue lines denote the envelope of the PI2018 samples; red dots are samples generated with BayOp.  These results are based on a total of 1600 (linear oscillation), 1400 (logarithmic oscillation) and 7300 (running logarithmic oscillation, including 2700 refinement samples (see text)) samples, respectively.  Since BayOp does not reject any sampled function values (unlike MCMC-based samplers), these numbers are equal to the number of times the likelihood was evaluated.}  
\end{figure}

In figure~\ref{fig:Planck_xisq}, we show the profile likelihood from the Pl2018 analysis in blue along with the samples generated by the BayOp analysis in red.  We find an excellent agreement between the two approaches for lower frequency features $\omega_X \lesssim 30$, indicating that fixing the \LCDM parameters in our analysis does indeed not affect the outcome.  At the higher frequencies, it becomes increasingly evident that the PI2018 results do not fully resolve the peak structure, particularly for the running logarithmic oscillation model (which has one more parameter than the other two models).  While PI2018 does identify most of the minor peaks at the correct frequencies, their amplitudes are systematically underestimated.

\begin{figure}[hpt]
    \centering
    \includegraphics[width=0.98\textwidth,height=510pt]{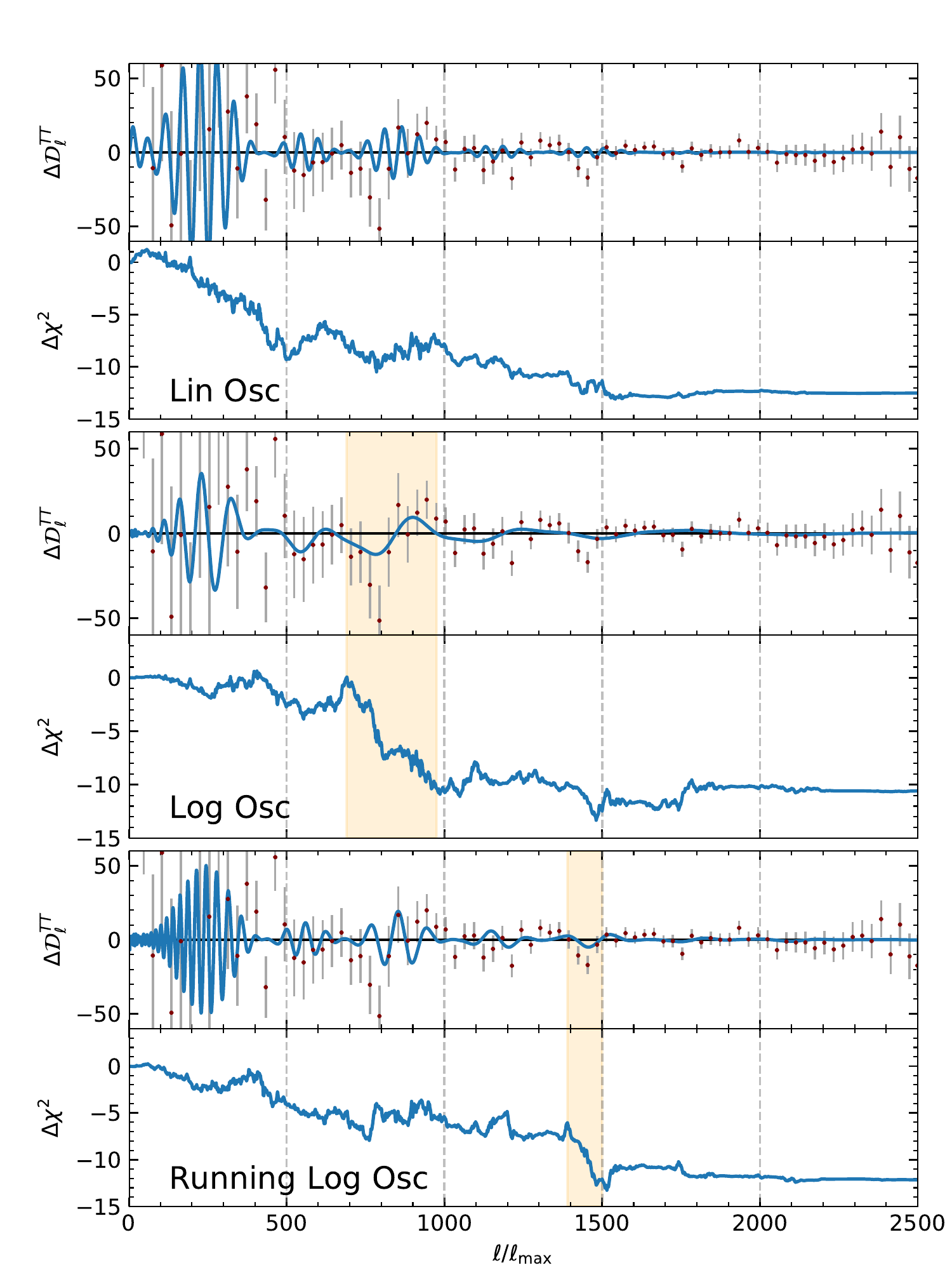}
    \caption{At the top of each panel we plot the residuals of the binned \Planck 2018 CMB temperature data (red) for the \LCDM best fit; the blue lines represent the best-fits of the linear, logarithmic and running logarithmic feature models (see table~\ref{tab:chisq} for the values of the best-fit parameters). At the bottom of each panel, we plot the cumulative improvement $\Delta\chi^2$ with respect to temperature+polarisation data as a function of $\ell_\mathrm{max}$. Large negative gradients in $\Delta\chi^2$ indicate regions of the spectrum where the features model does particularly well; we highlight two of these regions in orange. 
    \label{fig:Planck_Cl}}
\end{figure}

Notably, as can be seen from the top and bottom panels of figure~\ref{fig:Planck_Cl}, the best fits for the linear and running logarithmic oscillation models identified by the PI2018 analysis are not actually the global maxima, and even better fits can be achieved at very different frequencies.  For instance in the linear model, the global maximum at $\omega_{\rm lin} \approx 80$ yields an extra improvement of $\Delta\chi^2 \sim -2$ compared to the best fit found in PI2018.

This behaviour is likely due to a tendency of the nested sampling algorithm to not spend a lot of effort exploiting what it has identified as narrow local maxima that contribute little to the overall volume of the posterior.  As mentioned in section~\ref{sec:difference_PI2018} above, this is compounded by the logarithmic scaling of $\omega_X$ used by PI2018, which, compared to a linear scaling, ``squeezes'' high-frequency peaks relative to ones at lower frequencies.

The attentive reader will have noticed that the bottom panel of figure~\ref{fig:Planck_Cl} has significantly more samples than the other two panels.  For one, this is due to the higher dimensionality of this model.  Another reason is that in all models with four or more feature parameters, we ran a second round of BayOp on a more finely sampled grid on a narrow region of parameter space centered around the respective maxima in order to improve the estimate of the best fit (see appendix~\ref{app:refine} for details).  The total number of samples taken in the course of this refinement process is somewhat smaller than those taken in the first iteration, but the majority of them naturally yield a good fit (i.e., $\Delta \chi^2 < 0$) and are thus over-represented in the plot, since samples with $\Delta \chi^2 < 0$ are not plotted.

It is interesting to ask which part of the data is responsible for the improvement in the fit.  Do the primordial features actually correspond to localised features in the angular power spectrum $\mathcal{D}_\ell$?  To answer this question, we plot the residuals of the \Planck temperature data along with the cumulative $\Delta \chi^2$ (for the temperature+polarisation data) as a function of the maximum multipole $\ell_\mathrm{max}$ in figure~\ref{fig:Planck_Cl}.\footnote{Note that strictly speaking the correspondence between residuals and cumulative $\Delta \chi^2$ in figure~\ref{fig:Planck_Cl} is not quite exact since for reasons of clarity we only plot the binned temperature data, while the $\Delta \chi^2$ are calculated for unbinned temperature+polarisation data.  However, at all but the lowest multipoles, \Planck sensitivity to features is dominated by temperature data.} In all three models' best-fit spectra, data at multipoles $\ell \gtrsim 1700$ do not actually contribute much to the overall $\Delta \chi^2$.  That is not a surprising result, since the \Planck temperature power spectrum data start becoming noise-dominated for multipoles beyond $\ell \simeq 1600$~\cite{Planck:2018nkj}, which drastically reduces their sensitivity to global features.

The improvement in the linear oscillation model grows at a fairly steady rate as $\ell_\mathrm{max}$ is increased to about 1500, and no particular feature in the data can be identified to be responsible.  For the logarithmic oscillation and running logarithmic oscillation model, on the other hand, a large part of the total improvement can be attributed to two specific features in the \Planck temperature power spectrum.  The former traces the dip-bump structure around $\ell \sim 700 \hbox{-} 1000$, and the latter reproduces the dip around $\ell \sim 1450$ quite well.

\subsection{Primordial Standard Clock}
For a continuously ticking primordial standard clock, and ignoring a model-dependent envelope function, the primordial power spectrum is given by
\begin{equation} \label{eq:pscspectrum}
    \mathcal{P}^\mathrm{psc}_\mathcal{R}=\mathcal{P}^0_\mathcal{R}(k) \left[1+\mathcal{A}_{\rm psc} \cos \left(p\,\omega_{\rm psc} \left(\frac{k}{k_r}\right)^\frac{1}{p}+\varphi_{\rm psc}\right)\right]\,.
\end{equation}
This parameterisation actually covers a number of different physical settings depending on the parameter $p$.  Given that $a(t) \propto t^p$ during the generation of perturbations, we retrieve inflation if $p \gg 1$, a matter bounce for $p \approx 2/3$, and an ekpyrotic scenario for $0 < p \ll 1$, where the latter two correspond to a contracting universe at the time the CMB scales leave the horizon. 
In the following, we will consider these three cases separately.

If we assume that the primordial clock is set in motion at a time $t_r$ (e.g., by a turn in field space), and $k_r = H(t_r)$ is the wave number corresponding to the horizon size at that time, then modulations of the power spectrum will only occur at $k < k_r$ in expanding and $k > k_r$ in contracting models, respectively.  To allow for this scenario, we only use eq.~\eqref{eq:pscspectrum} for $k \lessgtr k_r$ (if $p \gtrless 1$), and set $\mathcal{P}^X_\mathcal{R}=\mathcal{P}^0_\mathcal{R}$ otherwise~\cite{Chen:2014cwa}.  The parameter $k_r$ therefore marks a transition from unmodulated to modulated power spectrum, and can be chosen such that only parts of the wavenumbers accessible to CMB data are affected by the modulation.  This opens the possibility to fit certain features in the data more aggressively without spoiling the fit in other parts of the power spectrum, which explains why the best fit modulation amplitudes for the inflationary and ekpyrotic PSC models are much larger than their counterparts in the models with global oscillations (see table~\ref{tab:chisq}).   Note that in all of these scenarios, the likelihood tends to also be multimodal in the $k_r$-direction (in addition to the multimodality in the frequency parameter direction), which poses a challenge to any sampling algorithm.

In order to get a rough idea which parts of the CMB angular power spectrum are subject to oscillations, we can relate $k_r$ to a transition multipole $\ell_r$ by approximating the square of the transfer function with a delta function centred on its maximum, resulting in an approximate relation $\ell_r\approx k_r/(7\times10^{-5})$ for the temperature power spectrum.  In the expanding scenario, features will accordingly occur at $\ell \gtrsim \ell_r$ and in the contracting scenarios at $\ell \lesssim \ell_r$.

\begin{figure}[hpt]
    \centering
    \includegraphics[width=0.98\textwidth,height=500pt]{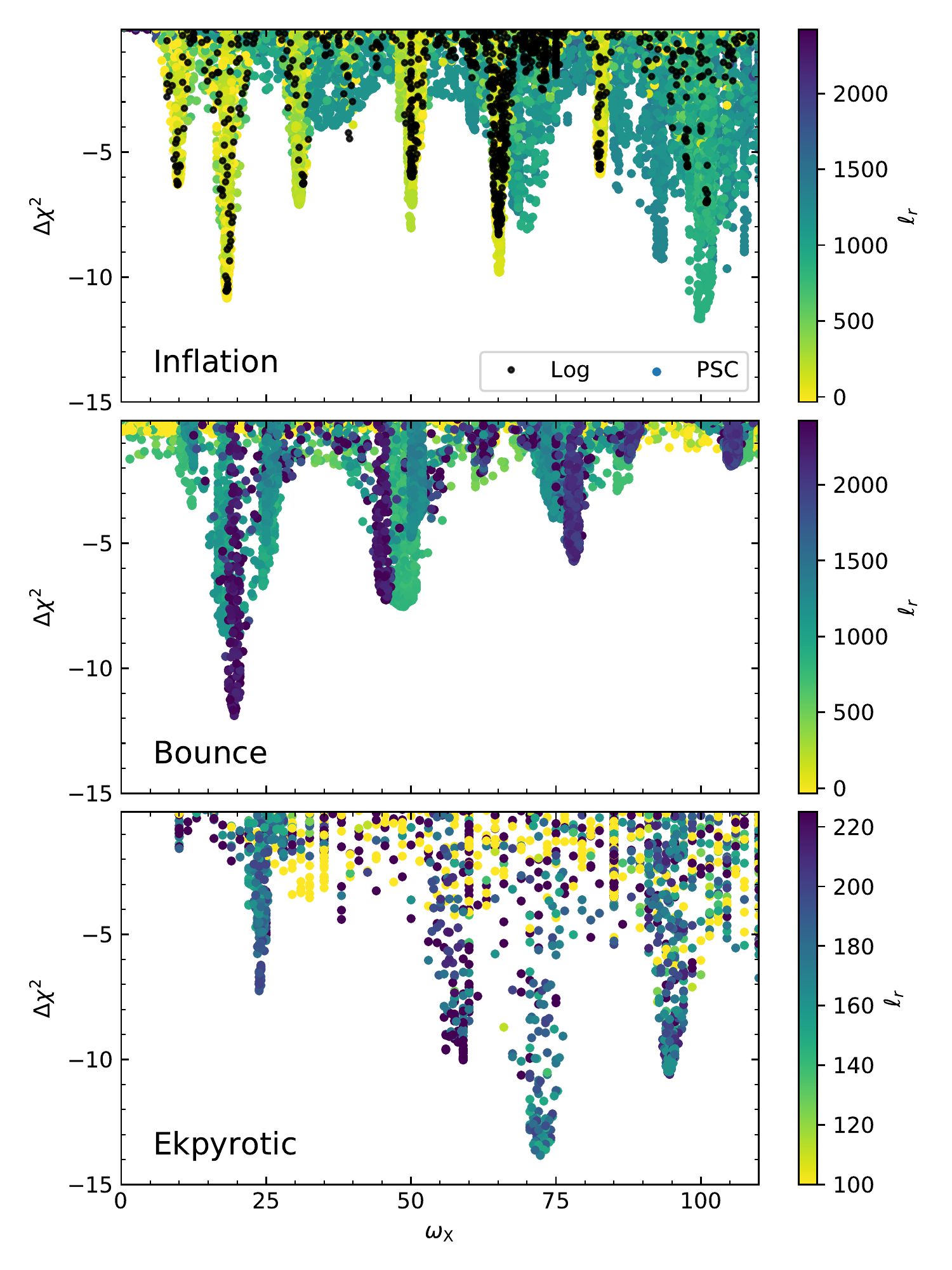}
    \caption{Improvement $\Delta \chi^2$ versus frequency parameter $\omega_{\rm X}$ for samples generated by BayOp for the different PSC models.  In the top panel, we plot the frequency $\omega_{\rm psc}$. In the centre and bottom panels, we show the effective frequency $\omega_{\rm eff}$ (eq. \eqref{eq:omega_eff}) with $k_p=0.11$ and $k_p=0.012$, respectively. For the inflationary PSC, we also display the results of the global logarithmic oscillations for comparison. The multipole $\ell_r$ describes the onset of the oscillation of the PSC. For inflation (top), the features appear only for $\ell \gtrsim \ell_r$, for the other two models for $\ell \lesssim \ell_r$. Note the different $\ell_r$ scale for the ekpyrotic model.
    \label{fig:PSC_xisq}}
\end{figure}

\begin{figure}[hpt]
    \centering
    \includegraphics[width=0.98\textwidth,height=520pt]{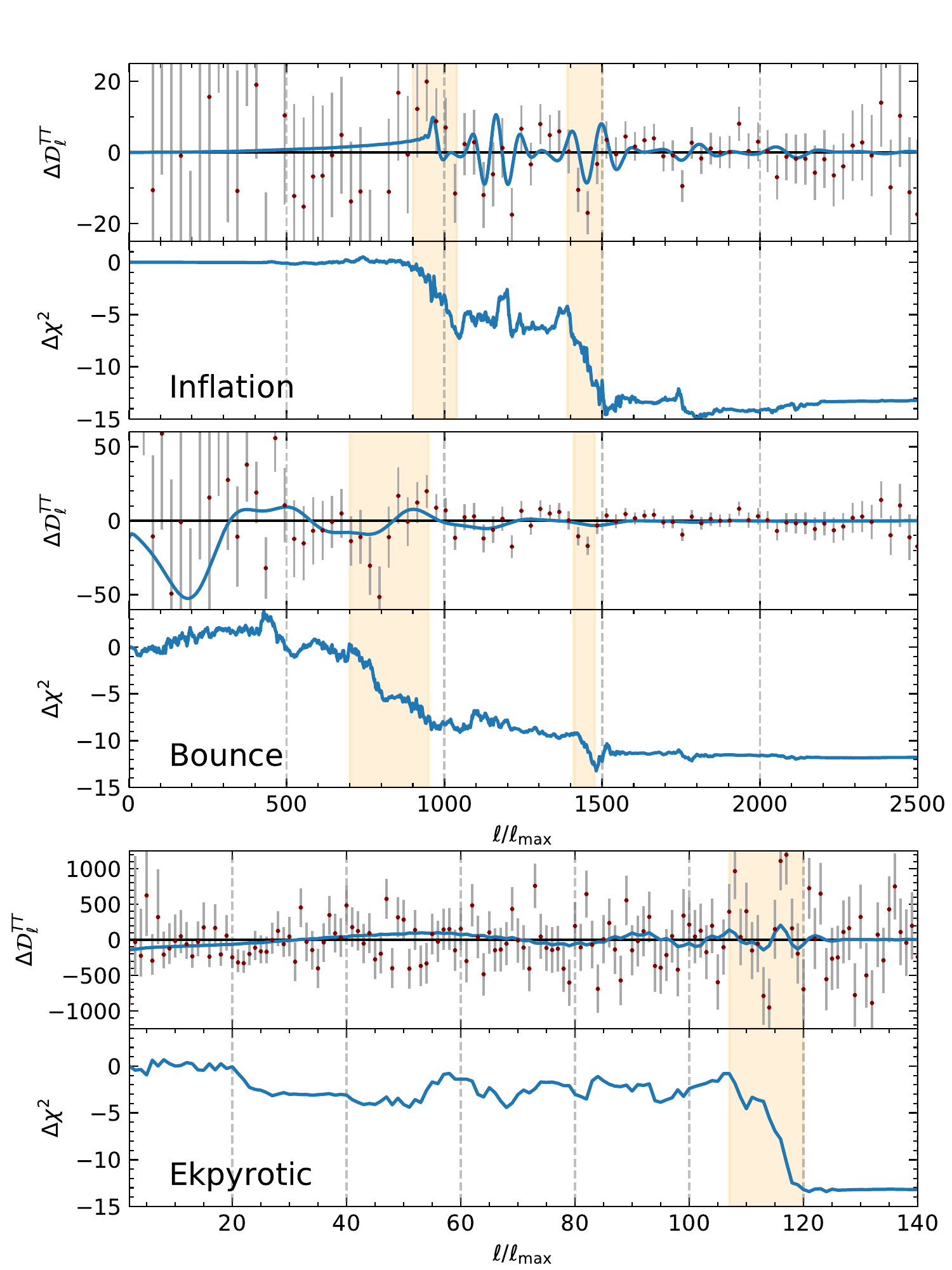}
    \caption{Residuals of the \Planck 2018 CMB TT data (red) with respect to the \LCDM best fit compared with the prediction of the PSC feature model best-fits (blue), and corresponding cumulative improvement $\Delta\chi^2$ with respect to temperature+polarisation data as a function of $\ell_\mathrm{max}$.  In the first two panels, the TT data are binned. The bottom panel has different scales in both $\ell$ and $\Delta\mathcal{D}^{TT}_\ell$ and shows the unbinned TT data.  Regions that contribute particularly significantly to the total improvement are highlighted in orange.
    \label{fig:PSC_CL}}
\end{figure}

\subsubsection{Inflation}
In the case of fast expansion with $p \gg 1$, and since $\lim_{p\to \infty} \tfrac{\mathrm{d}}{\mathrm{d}x} \left( p \, x^{1/p} \right) = \tfrac{\mathrm{d}}{\mathrm{d}x} \left( \ln x \right)$, the power spectrum in eq.~\eqref{eq:pscspectrum} can be approximated by
\begin{equation}\label{eq:pscinf}
    \mathcal{P}^\mathrm{psc,inf}_\mathcal{R}\approx\mathcal{P}^0_\mathcal{R}(k) \left[1+\mathcal{A}_{\rm psc} \cos \left(\,\omega_{\rm psc}\, \ln\left(\frac{k}{k_r}\right)+\varphi_{\rm psc}\right)\right]\,,
\end{equation}
where the constant offset in the argument of the cosine was absorbed in a redefinition of the phase.  For $k > k_r$ this expression is identical to the global logarithmic oscillation case of eq.~\eqref{eq:parametrisation}, modulo another redefinition of the phase.  This allows for direct comparison with the results found in the previous section.

In the top panel of figure~\ref{fig:PSC_xisq}, we show the $\Delta \chi^2$ of BayOp samples for the inflationary PSC scenario and for the global logarithmic oscillation model. For the PSC samples, the colour of the dots indicates the multipole which corresponds to the onset of oscillations in the CMB temperature spectrum, $\ell_r$. 
We split the prior range for $\ell_r$ into two intervals for the analysis: $\ell_r<1250$ and $\ell_r>1250$, but combine the samples for the plot.
As expected, the small $\ell_r$-interval (yellow to yellow-green dots) of the inflationary PSC and the global logarithmic models leads to peaks appearing around the same frequencies, since for $\ell_r \rightarrow 0$, the two signals are identical.  Slight improvements over the global logarithmic modulation can be achieved if the start of oscillations is tuned in such a way that the large scales ($\ell<250$) are not subject to modulations.

For larger values of $\ell_r$, i.e., if the spectrum is only modulated at small scales, we observe a whole range of new peaks at frequencies that do not provide a good fit in the global oscillation case.  With these features being more localised, they also allow for a larger modulation amplitude.  In fact, the global maximum of the inflationary PSC model falls into this category, with $\ell_r \approx 986$ and $\mathcal{A}_\mathrm{psc} \approx 0.07$, yielding $\Delta \chi^2 = -13.2$, an improvement of $2.6$ over the global oscillation model (at the cost of one extra parameter though).  As can be seen from the top panel of figure~\ref{fig:PSC_xisq}, this is achieved primarily by an excellent fit to the bump around $\ell \sim 900$ and the dip around $\ell \sim 1400$ in the \Planck temperature power spectrum.

Let us also mention a recent analysis of the inflationary PSC scenario in ref.~\cite{Braglia:2021ckn}, where the primordial power spectrum was directly calculated from the level of the Lagrangian using the  \texttt{BINGO}~\cite{Hazra:2012yn} code, and parameter space was explored with the \texttt{PolyChord} sampler~\cite{Handley:2015fda}.  The resulting power spectra qualitatively differ from the ones in this work in that they naturally incorporate a $k$-dependent envelope for the oscillations which is not modelled in the approximation of eq.~\eqref{eq:pscinf}.  This makes a straightforward comparison of results difficult, but we do note the presence of a major peak around values of the frequency parameter $\omega_\mathrm{psc} \sim 20$ in both cases.

\subsubsection{Matter bounce}
Inflation is not the only way for a minimally coupled massless field to produce a nearly scale-invariant spectrum of perturbations on superhorizon scales.  It was shown in ref.~\cite{Wands_1999} that the spectral index of the field's perturbations satisfies
\begin{equation}\label{eq:ns_bounce}
n_\mathrm{s} - 1 = 3-\left| 3+\frac{2}{p-1}\right|\,
\end{equation}
if the background evolves as $a(t) \propto t^p$.  So a matter-dominated universe with $p \sim 2/3$ can result in near scale-invariant perturbations as well, but in order for fluctuations to exit the horizon, such a universe must of course be contracting at that time (and subsequently undergo a bounce).  This scenario is referred to as a \textit{matter bounce} and solves the horizon and flatness problems of Standard Big Bang cosmology just like inflation does (see, e.g., ref.~\cite{Brandenberger:2012zb} for a review of contracting primordial scenarios).

In the contracting case, modulations to the primordial power spectrum appear for $k<k_r$ because the horizon contracts faster than the scales $k$ \cite{Chen:2014cwa}. At the onset of the modulations, the scale $k_r$ exits the horizon and scales exiting after that have $k<k_r$.
This is opposite to the inflationary case where the scales $k$ expand faster than the horizon.
Therefore, the CMB power spectrum becomes unmodulated for $\ell \gtrsim \ell_r$, opposite to the inflationary example above. Thus, this model can achieve a better local fit to features at smaller $\ell$ without needing to worry about ruining the fit at large multipoles.

The contracting scenarios exhibit a strong degeneracy between between $p$, $k_r$ and the frequency parameter $\omega_\mathrm{psc}$.  In order to lift the degeneracy it is convenient to define an effective frequency parameter
\begin{equation}\label{eq:omega_eff}
    \omega_{\rm eff}\equiv p\,\omega_{\rm psc}\left(\frac{k_p}{k_r}\right)^\frac{1}{p}\,.
\end{equation}
with a pivot scale $k_p$.  In terms of the effective frequency the primordial power spectrum for $k < k_r$ can then be written as
\begin{equation}\label{eq:PSCcontracting}
    \mathcal{P}^{\rm psc,contr}_\mathcal{R}(k)=\mathcal{P}^0_\mathcal{R}(k)\left[1+\mathcal{A}_{\rm psc} \cos\left(\omega_{\rm eff} \left(\frac{k}{k_p}\right)^{\frac{1}{p}}+\varphi_\mathrm{psc}\right)\right]\,.
\end{equation}
Compared to the global linear oscillation model, the power spectrum in this model not only exhibits a transition from modulated to unmodulated at $k_r$  but also a scale-dependent modulation frequency, whose running is determined by $p$.  
Since in this scenario the parameter $p$ also sets the spectral index of the underlying power spectrum via eq.~\eqref{eq:ns_bounce}, we fix $p=0.668$ to recover the \Planck best-fit value of $n_\mathrm{s}$. The pivot scale is taken to be $k_p = 0.11~\mathrm{Mpc}^{-1}$ such that the effective frequency parameter of this model matches the frequency of the global linear oscillations model around $k_*$.

In the middle panel of figure~\ref{fig:PSC_xisq}, we plot the effective frequency parameter versus $\Delta\chi^2$. As in the previous subsection, the  analysis was split into a low- and high-$\ell$ part.  It turns out the data prefer the modulations to extend over nearly the entire range of $\ell$ up to 1500, i.e., the onset of oscillations should preferably occur at wavelengths corresponding to $\ell_r > 1500$. The global maximum around $\omega_{\rm eff} \approx 20$ yields an improvement of $\Delta\chi^2\approx -12$, and can be achieved for wide range of $k_r-\omega_{\rm psc}$ combinations.  Inspection of the top panel of figure~\ref{fig:Planck_xisq} reveals a local maximum of the posterior at the same value of the (constant) frequency parameter for the linear oscillation model; in the matter bounce model, the running of the effective frequency parameter leads to an enhancement of this peak.

As illustrated in the middle panel of figure~\ref{fig:PSC_CL} and already observed in the global oscillation models, there is not much contribution to the overall $\Delta \chi^2$ from multipoles $\ell \gtrsim 1500$; the best-fit primordial features are essentially invisible to the \Planck data in this range.

\subsubsection{Ekpyrosis}
As a third PSC scenario, we study the case of features generated during a phase of slow contraction with $0 < p \ll 1$.  If the underlying unmodulated power spectrum is to be nearly scale-invariant, the usual generation mechanism of adiabatic perturbations that works well for inflation or a matter bounce cannot be applied in this case and the spectral tilt $n_s$ is not described by eq.~\eqref{eq:ns_bounce}.  Instead, the solution found in ekpyrotic cosmologies~\cite{Lehners:2008vx} is to first generate isocurvature perturbations which are later converted into curvature perturbations. A slight red tilt, as favoured by observations stems from a decrease in the steepness of the potential of the scalar field dominating the energy density of the universe during the ekpyrotic epoch~\cite{Lehners:2007ac}.

Ekpyrotic models typically span a range of $0<p<1/3$~ \cite{Ijjas:2013sua}.  However, since values of $p$ close to zero lead to extremely strong running of the frequency in eq.~\eqref{eq:PSCcontracting} with the frequency parameter very quickly becoming too large to be resolved by the data, we restrict the prior range to $p\in[0.2, 0.3]$.
Additionally, we narrow down the search for features to the phenomenologically most interesting small multipole region $\ell < 220$
and use $k_p=0.012~\rm{Mpc}^{-1}$ for the effective frequency (eq.~\eqref{eq:omega_eff}), which matches the frequency parameter of the global linear oscillation model around wavenumbers corresponding to $\ell \sim 100$.

The best fit for this model has $\ell_r \sim 110$ and $p\approx0.24$, with a rather large modulation amplitude of about $30\%$. We show the residuals and the cumulative $\Delta \chi^2$ in the bottom panel of figure~\ref{fig:PSC_xisq}.  Interestingly, most of the improvement is obtained over a narrow range of about $\Delta \ell \sim 10$ around $\ell=108-120$ (highlighted in the Figure) by providing a better match than a smooth power spectrum to the two pairs of outlying data points in this region.  

\subsection{Profile likelihoods}\label{sec:profile}
Even though BayOp is technically intended to only find the global maximum of a function, the information gained about the function in the course of the sampling process eventually also results in an estimate of the overall shape of the function as well as the uncertainty of this estimate, both via the application of GPR.

\begin{figure}[htp]
\centering
    \includegraphics[width=0.98\textwidth]{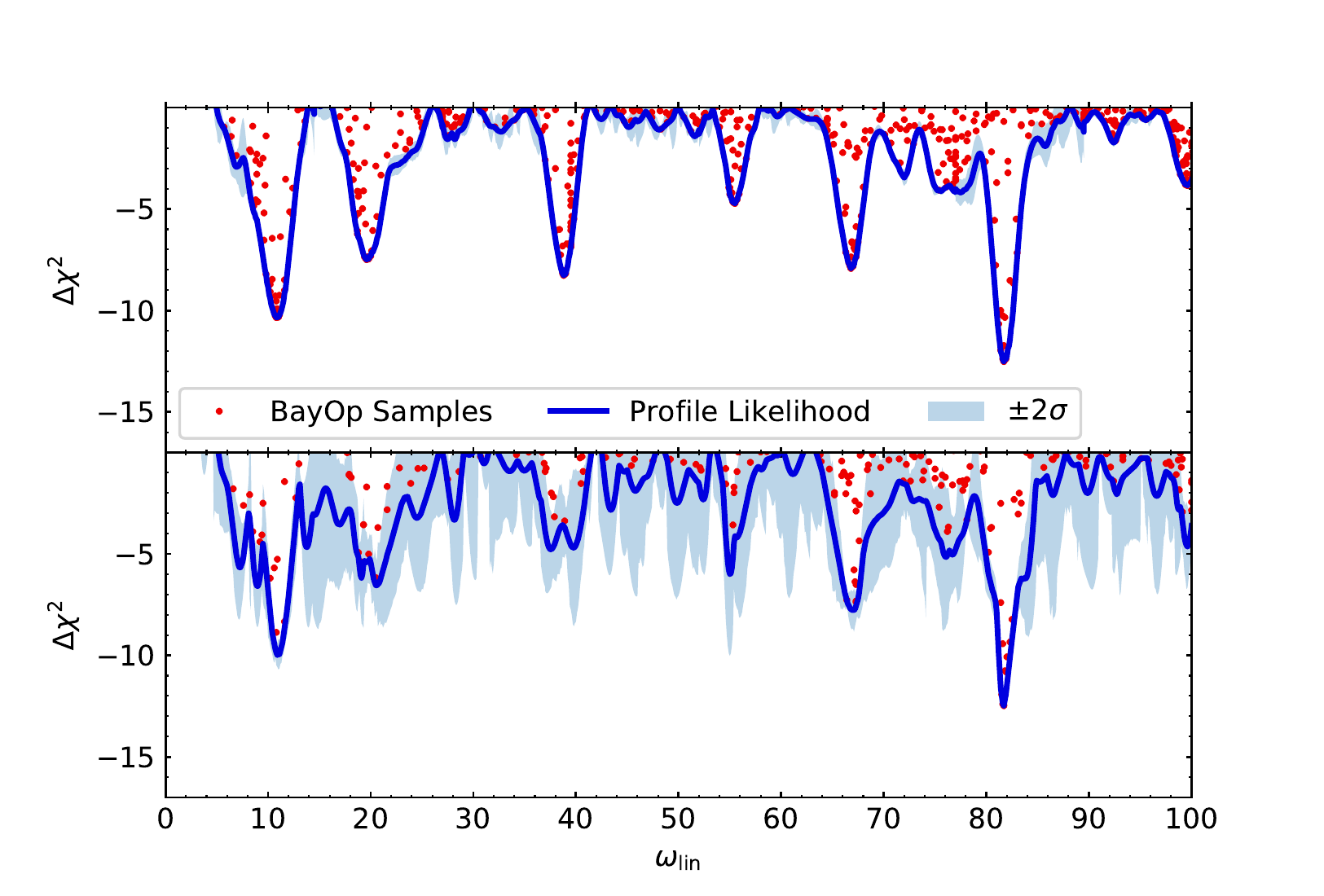}
    \caption{Profile likelihood of the linear oscillations model frequency parameter and associated $2\sigma$-uncertainty band, calculated with BayOp. \textit{Top:} results from an analysis with the frequency parameter range split into eight separate bins (about 1700 samples taken in total). \textit{Bottom:} results without binning ($\sim 600$ samples).}
    \label{fig:profile_freq}
\end{figure}

As already shown above, one can for instance extract profile likelihoods from the GPR results by identifying the maximum of the GPR mean $\mu$ on slices of the regression grid with a given fixed value of whichever parameter is of interest (e.g., the modulation frequency).  Similarly, we can construct, e.g., 2-$\sigma$ uncertainty intervals by finding the maxima of $\mu \pm 2 \sigma$ on the same slices, where $\sigma$ is the square root of the GPR variance.  Neither of these operations requires a large additional computational effort.
Here we explicitly demonstrate that even for a relatively modest number of samples and not just in parameter regions close to the global maximum, our implementation of BayOp leads to serviceable likelihood profiles, with remaining uncertainties quantified by the covariance of the GPR.
As an example, we plot our estimates of the profile likelihood for the frequency parameter direction of the (3-parameter) global linear oscillation model in figure~\ref{fig:profile_freq}, along with the corresponding uncertainty.  In the top panel, the frequency parameter range was split into 8 separate sections (bins) of equal width before applying the BayOp algorithm to each of them individually, resulting in a total of $\sim 1700$ samples.  The 2$\sigma$-uncertainties are very small indeed near the peaks and reach at most $\mathcal{O}(1)$ in lower-likelihood regions.  For comparison, we include in the bottom panel the case where the entire frequency range was analysed without binning.  This estimate is based on considerably fewer samples ($\sim 600$)\footnote{In comparison, the PI2018 results for this model are based on approximately 50,000 samples with an acceptance rate of roughly 0.3 requiring about 180,000 total likelihood evaluations for a 9-dimensional parameter space. Using the scaling of required samples and acceptance rate with dimensionality in table 3 of ref. \cite{Feroz:2008xx}, we estimate that in the 3-dimensional case \texttt{MultiNest} would have required of order 25,000 likelihood evaluations, i.e., over an order of magnitude more than BayOp.}, and therefore naturally exhibits much larger uncertainty away from the major peaks.

\begin{figure}[htp]
    \centering
    \includegraphics[width=0.98 \textwidth]{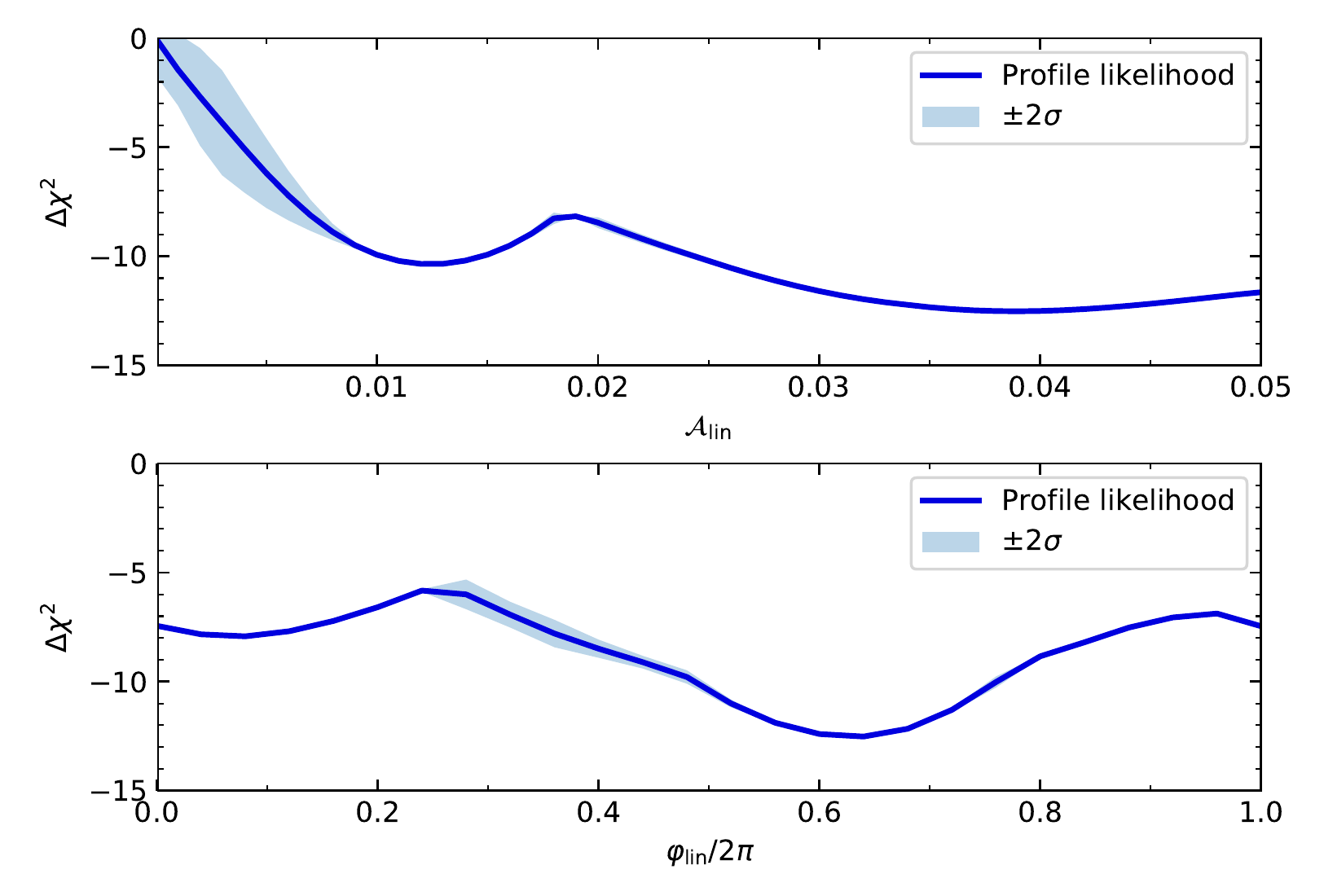}
    \caption{Profile likelihoods and $2\sigma$-uncertainty bands for the amplitude (\textit{top}) and phase (\textit{bottom}) parameters of the 
    global linear oscillation model. In both cases, the frequency prior was split into eight bins.
    \label{fig:profile_ap}}
\end{figure}

While the frequency parameter direction is clearly the most interesting one due to its complex structure, one can just as well construct the corresponding profile likelihoods in the other parameter directions from our BayOp samples.  We show these results for the linear oscillations model in figure~\ref{fig:profile_ap}.  Note that it would be very difficult to obtain trustworthy results at all if one were to use the conventional method for calculating profiles, i.e., via application of an optimisation algorithm to a grid of fixed parameter values.  Also, one should keep in mind that our implementation of BayOp is not optimised for the construction of profile likelihoods -- they are merely a convenient by-product of our goal of global likelihood optimisation.  A different choice of acquisition function that puts more emphasis on exploring regions with high uncertainty rather than exploitation would plausibly give comparable results to ours with considerably fewer function evaluations.

In addition to using the results of BayOp for evaluating profile likelihoods by maximising over other parameter directions, it is also conceivable to \textit{integrate} instead, provided a prior probability density has been specified for parameter space.  This would result in marginal Bayesian posterior probabilities or, if one integrates over all parameter directions, even estimates of the Bayesian evidence and its uncertainty.  We will leave the exploration of this intriguing possibility to future work.

\section{Conclusions \label{sec:conclusions}}

In this work, we have applied a new sampling algorithm based on Bayesian Optimisation to the analysis of \Planck CMB data for a range of inflation models with oscillatory features.  These models are notoriously tricky to deal with due to their tendency to exhibit highly complex likelihood functions that are difficult to explore, and due to the relatively long time required for a single likelihood evaluation.

We have demonstrated the excellent efficiency and performance of our approach by direct comparison with the results of previous investigations based on Monte Carlo sampling methods~\cite{Akrami:2018odb}, discovering new, previously missed global likelihood maxima for the linear and running logarithmic oscillation models which improve the fit to the data by an additional $\Delta \chi^2$ of up to 2.  For all models considered, the optimal parameter values improve the fit
by \mbox{$-\Delta\chi^2\sim10$--$14$} with respect to a power-law primordial spectrum, at the cost of having to introduce 3--5 extra parameters.  

However, as shown in ref.~\cite{Planck:2015sxf}, in oscillatory features models and for simulated \textit{Planck}-like data, a $\Delta\chi^2$ of this order is not incompatible with a ``no features'' null hypothesis, so our results do not imply any statistically significant discovery of features.  Nonetheless, upcoming high-resolution CMB polarisation measurements such as CMB-S4~\cite{Abazajian:2019eic} will give us an independent handle on features, and if any of the current promising feature candidates are real, ought to contribute enough additional constraining power to push their evidence from marginal to significant~\cite{Miranda:2014fwa}.

Overall, we believe that the BayOp method introduced in this work represents a very useful addition to a feature-hunters arsenal and can generally serve as a complement to more traditional sampling methods for the analysis of cosmological data.  We look forward to exploring the application of BayOp to constructing marginalised posteriors and estimating Bayesian evidences in an upcoming paper.

\appendix
\section{Optimising the optimisation \label{app:A}}
In the Appendix, we present several tweaks to the algorithm which serve to speed up the analysis or increase the quality of the final result.

\subsection{Number of random samples for initialisation \label{app:random}}
While the algorithm could also be started from an empty set of data points, this would often result in the GPR initially returning an estimate of the function as slowly varying with large uncertainties, which might in practice lead to issues with numerical stability and drawing samples mostly at the edges of parameter space.  This issue can largely be avoided by drawing a number of samples randomly from the prior.  If one wants to get the most out of the advantages of BayOp, one should of course strive to keep this number small.

Simpler functions with little structure would obviously require fewer initial samples than complex ones. For the examples considered in this paper, we found that taking 20-40 initial random samples per parameter direction is sufficient to avoid these issues.  In our analysis, we used 100 initial samples for the 3-parameter models and 150 initial samples for the 4- and 5-parameter models. 

\subsection{Removing points from the grid \label{app:remove}}
The step of predicting the function via GPR requires the solution of a system of $n_{\rm data} \times n_{\rm grid}$ linear equations, which has a complexity of $\mathcal{O}(n_{\rm data}^2 n_{\rm grid})$ for $n_{\rm grid}>n_{\rm data}$.  With $n_{\rm grid} \gg n_{\rm data}$ initially and $n_{\rm data}$ constantly growing, the time required for this step will thus roughly increase like $n_{\rm data}^2$ and therefore eventually take up a larger and larger fraction of the computing time of each BayOp iteration.  

However, there is a way to counterbalance this growth, namely by reducing $n_{\rm grid}$.  Once the function is reasonably well known, grid points with expected improvement below a threshold value ($\mathrm{EI} < \epsilon_{\rm EI}$) may no longer be worth considering and can be removed from the grid.  Such points are typically in close vicinity of samples that returned a poor value compared to the best-observed value.

We start with a very conservative threshold ($\epsilon_{\rm EI}=10^{-50}$) because for low $n_{\rm data}$, the prediction of the underlying function (and hence the EI) can still change significantly.  
With more samples generated and the prediction getting more and more robust, we gradually increase the threshold until we arrive at $\epsilon_{\rm EI}=10^{-5}$. In this way, we avoid that the computational costs of the GPR step contribute significantly to the total run time.  With our settings, the contributions stay insignificant ($\lesssim 1\%$) at all times.

The choice of the final threshold does not impact the runtime significantly. In most cases, the maximum value of the EI drops very quickly once all relevant local maxima have been discovered and exploited. For the 3-parameter models, the maximum of the EI typically drops from $10^{-2}$ to $10^{-10}$ within 10-20 BayOp iterations.

\subsection{Refinement sampling around the maximum \label{app:refine}}
Instead of removing points from the grid, one can just as well add points to the grid at any time.  This can be a useful strategy in case one would like to better resolve the function in particularly interesting regions of parameter space, similar in spirit to, e.g., adaptive mesh refinement techniques in field theory simulations.

Especially in higher-dimensional problems, the exponential scaling of $n_\mathrm{grid}$ with the number of dimensions implies strict limits to the achievable resolution for a given amount of computer memory, necessitating an initially rather coarsely-spaced grid.  
Once the coarse structure of the function is learnt and the approximate location of its maximum has been found, one might therefore want to add more finely spaced grid points to increase the resolution around the coarse grid maximum in order to determine the global function maximum more accurately.  

We implement a simple refinement sampling strategy along these lines in all of our 4- and 5-parameter models by effectively running BayOp twice.  First, it is run on a relatively coarse grid covering all of parameter space (or the entire bin, see appendix~\ref{app:split} below).  Then, a second, finer grid with 5 times higher resolution is centred around the location of the coarse grid's best fit point and BayOp is restarted, including all the data points taken in the first run.

\subsection{Updating the hyperparameters less frequently \label{app:bobyqa}}
Finding the optimal choice of hyperparameters given the data requires solving a \mbox{$n_{\rm data}\times n_{\rm data}$} linear system multiple times to find the maximum of eq.~\eqref{eq:MLE}. The computational complexity for this process scales like $\mathcal{O}(n_{\rm data}^3)$.

While initially, this step will contribute only a negligible fraction of the computation of a BayOp step, this changes as more samples are collected, and for $n_{\rm data} \gtrsim 500$ it becomes the most time-consuming part of the calculation.  Fortunately, at that stage, the function is already reasonably well known and the prediction is quite robust, such that the optimal values of the hyperparameters no longer change significantly at each step.  In the examples, we considered, the total change of the hyperparameters over 100 iterations of BayOp between steps 500 and 600 was only a few percent.  Once a certain number of data points is reached, it is therefore no longer necessary to update the hyperparameters every single step, and we gradually reduce the frequency of hyperparameter updates to once per 100 steps.

\subsection{Parallelisation by splitting parameter directions \label{app:split}}
The execution of the code can be trivially parallelised by dividing parameter space into multiple disconnected parts (bins).  Besides allowing in principle for a higher density of target points in each bin, binning comes with a couple of other advantages.

Firstly, the optimal choice of hyperparameters may be parameter dependent, e.g., the optimal prior width tends to decrease with the modulation frequency.  Splitting the frequency direction into several bins allows us to take into account the variation of hyperparameters to some extent, resulting ultimately in a more precise GPR and thus a better prediction of the likelihood.

Secondly, not only the global maximum but also the local maximum within each bin will be exploited thoroughly, which leads to a better resolution of the whole function and higher quality profile likelihoods, as demonstrated in section~\ref{sec:profile}.

Our tests of the impact of binning on the cumulative run time of the code with fixed target point density showed a slight increase with the number of bins due to the initial overhead of random samples having to be taken for each bin. However, the total wall time decreases as a function of the number of bins, since smaller bins are explored more quickly.

For the analysis in this paper, we opted to split the frequency direction into 8 sections. This strikes a good balance between run time and resolving  the structure of the likelihood. Our analysis of the efficiency as a function of bins shows that a good choice of bin width is given by around 4 correlation lengths. This is plausible since the correlation of a point to its surrounding essentially vanishes at a distance of around 2-3 correlation lengths. For the inflationary and bounce PSC models, we additionally divided the $k_r$-parameter direction into two sections to better resolve the local maxima, given that this direction is also multimodal, leading to 16 bins in total.  

With one CPU per bin, the 3-parameter models require a wall time of less than 5 hours and the 4-parameter models about 10-20 hours. The ekpyrotic model with 5 parameters was run on a small range of $k_r$ and took 40h to complete, using 8 bins.
All of these analyses can therefore realistically be run on desktop computers and do not explicitly require massive HPC resources.

\acknowledgments
Most of the numerical calculations for this work were performed on the computational cluster \textit{Katana}, supported by Research Technology Services at UNSW Sydney~\cite{Katana}.  We thank Steen Hannestad for discussions at an early stage of the work.

\bibliographystyle{apsrev}
\bibliography{biblio}

\end{document}